# Time-resolved neutron imaging at ANTARES cold neutron beamline

A.S. Tremsin[a,*], V. Dangendorf[b], K. Tittelmeier[b], B. Schillinger[c], M. Schulz[c], M. Lerche[d], W. B. Feller[e]

[a] *Space Sciences Laboratory, University of California at Berkeley, 7 Gauss Way, Berkeley, CA 94720, USA*

[b] *Physikalisch-Technische Bundesanstalt (PTB), Bundesallee 100, 38116 Braunschweig, Germany*

[c] *Technische Universität München, Heinz Maier-Leibnitz Zentrum (MLZ), Lichtenbergstr.1 D-85748, Garching, Germany*

[d] *University of California, Davis, McClellan Nuclear Research Center, 5335 Price Avenue, McClellan, CA 95652, USA*

[e] *NOVA Scientific, Inc., 10 Picker Rd., Sturbridge, MA 01566, USA*

E-mail: `ast@ssl.berkeley.edu`

ABSTRACT: In non-destructive evaluation with X-rays light elements embedded in dense, heavy (or high-Z) matrices show little contrast and their structural details can hardly be revealed. Neutron radiography, on the other hand, provides a solution for those cases, in particular for hydrogenous materials, owing to the large neutron scattering cross section of hydrogen and uncorrelated dependency of neutron cross section on the atomic number. The majority of neutron imaging experiments at the present time is conducted with static objects mainly due to the limited flux intensity of neutron beamline facilities and sometimes due to the limitations of the detectors. However, some applications require the studies of dynamic phenomena and can now be conducted at several high intensity beamlines such as the recently rebuilt ANTARES beam line at the FRM-II reactor. In this paper we demonstrate the capabilities of time resolved imaging for repetitive processes, where different phases of the process can be imaged simultaneously and integrated over multiple cycles. A fast MCP/Timepix neutron counting detector was used to image the water distribution within a model steam engine operating at 10 Hz frequency. Within <10 minutes integration the amount of water was measured as a function of cycle time with a sub-mm spatial resolution, thereby demonstrating the capabilities of time-resolved neutron radiography for the future applications. The neutron spectrum of the ANTARES beamline as well as transmission spectra of a Fe sample were also measured with the Time Of Flight (TOF) technique in combination with a high resolution beam chopper. The energy resolution of our setup was found to be ~0.8% at 5 meV and ~1.7% at 25 meV. The background level (most likely gammas and epithermal/fast neutrons) of the ANTARES beamline was also measured in our experiments and found to be on the scale of 1-2% when no filters are installed in the beam.

---

[*] Corresponding author.

**Contents**

# 1. Introduction

The high intensity of the recently recommissioned ANTARES cold neutron beamline facility [1] provides the possibility to study dynamic phenomena where the imaging acquisition times need to be relatively short in experiments where processes have typical time constants on a sub-second scale. The studies of periodic dynamic processes represent a subset of dynamic experiments, where multiple phases of the process can be imaged repetitively over multiple cycles in order to acquire sufficient neutron statistics. In such experiments the capability of the detection system to image individual phases of the process becomes as crucial as the intensity of the neutron beam. Stroboscopic imaging is one of the possible approaches, where a particular phase of the process is imaged for each cycle [2]-[5]. If only one phase of the process needs to be acquired, stroboscopic imaging can be performed with conventional gated detectors [6], [7], where neutrons arriving at a particular time interval after the trigger are registered and the rest of the neutrons are ignored. However, if multiple phases are to be studied the single phase stroboscopic imaging has to be repeated multiple times and the experiment time in such cases increases substantially. Also the compromise between the number of phases, measurement time and timing resolution has to be found for a particular process. The simultaneous acquisition of several phases of the process [8] is a very attractive alternative in dynamic imaging experiments and can be enabled by detection systems capable of acquisition of multiple frames per single cycle. Another approach is to detect both position and time for each registered neutron (event counting detectors) and thus acquire multiple phases simultaneously by sorting the neutrons according to their time of arrival at the detector. The counting rate capabilities of such neutron counting detectors usually are inferior to the frame-integrating devices and in some cases cannot utilize the full flux of existing bright neutron beamlines [1], [9], [10], [11].

　　　In this paper we report the results of time-resolved imaging performed with an MCP/Timepix detector [12], [13] operating in two different modes: (1) event counting mode with sub-µs time resolution for each detected neutron and (2) multiple images per trigger mode with timing resolution being the width of the frame, with 55 µm spatial resolution in both cases.

In mode (1) our existing electronics allows only one neutron per frame to be detected within a pixel (512x512 independently operating pixels), detecting ~4x10$^6$ n/cm$^2$/s at 10% overlaps. Multiple images per single trigger can still be acquired in Mode (1). Mode (2) allows detection of >10000 events/pix/frame with 255 independent images per single trigger, thus enabling simultaneous imaging of 255 phases of the process with 320 µs readout time between the frames. That mode (2) allows more efficient utilization of high neutron fluxes and is sufficient for processes with periods larger than ~10 ms, while faster processes can be studied with the event counting mode (1).

The dependence of the neutron flux on energy was measured for the ANTARES beamline. The energy of registered neutrons was determined by the Time Of Flight (TOF) method in combination with a beam chopper installed 7.41 m upstream from the detector. The spectrum of the open beam (no filters) as well as the transmission of a BCC steel sample were acquired simultaneously. In addition, the spectrum of the neutron beam with Beryllium and Lead filters was also measured, all with the detector mode (1).

The unique capability of neutron imaging to visualize the distribution of hydrogen containing substances (e.g. oil, petrol, water) within metal structures (such as engines) has been demonstrated previously [2]-[4], [14]. In Section 3.2 the dynamic study of water distribution within a model steam engine is presented, where multiple phases of the engine operating at 10 Hz frequency were acquired simultaneously. The high count rate mode (2) with 75 simultaneously imaged phases was implemented for the study of water distribution with 1 ms-wide frames, demonstrating the time-resolved imaging capabilities of the bright collimated neutron beamline in combination with a fast neutron counting detector.

## 2. Experimental setup and data acquisition

The neutron counting capability of the MCP/Timepix detector [12], [13] was utilized in our experiments performed at the ANTARES [1] neutron imaging beamline at the FRM II reactor at Technische Universität München. The detector was installed at ~16 m from the source. The active area of the detector is 28x28 mm$^2$ (512x512 pixels) determined by the 2x2 array of the Timepix readout chips [15]. A set of neutron sensitive Microchannel Plates (MCPs) with ~ 8 µm pores was converting incoming neutrons into a charge signal of ~105 electrons, all spatially confined within several pores, enabling imaging with the 55 µm resolution defined by the pixel size of the readout. The detector can operate in several modes as described in references [12], [13] and has a readout time of ~320 µs and can acquire ~1200 frames/sec. The counting rate of the detector can be as high as ~10$^8$ n/cm$^2$/s in event counting mode when the time resolution of the detector is determined by the width of the acquisition shutter. In the high timing resolution mode (< 1 µs) the event rate is limited to ~10$^6$ n/cm$^2$/s by the functionality of the current generation of Timepix readout to detect only one neutron per pixel per frame. The new generation of Timepix chips with sparcified readout [16] allows for 2x10$^7$ events/cm$^2$/s and will be implemented in our future detectors. For the time resolved imaging of cyclic processes an external trigger initiates the acquisition of up to 255 shutters with predefined times for each of them. The data from the Timepix readout was transferred over a 10 Gb connection through a 40 m fiberoptic cable to a PC where all events were sorted out in real time into the appropriate images corresponding to a particular phase of the process. The result of each time-resolved imaging experiment was a set of images, each corresponding to a given phase of the cyclic process.

## 2.1 Measurement of the spectra of ANTARES beamline

For the measurement of the beamline spectra a neutron chopper operating at ~ 3000 rpm (~50 Hz) with a ~3.2 mm pinhole positioned at ~175 mm radius was installed 7.41 m upstream of the detector in the beam. The chopper reduced the thermal and cold neutron flux to 0.35% of the full beam intensity, while the flux of neutrons with energies above 1 eV and gammas was almost not affected by the Cd disk of the chopper. The detector acquisition was synchronized to the chopper frequency of ~50 Hz. The timing of each neutron within the acquisition frame was registered with ~1 µs resolution relative to the chopper trigger. The upstream aperture of 9 mm in the collimator selector wheel was used in these experiments.

## 2.2 Time-resolved imaging of water distribution within steam engine

The dynamics of water distribution in the steam engine was measured with the fast acquisition mode (2). The 100 ms period of the engine operation was divided into 75 phases, each consisting of a 1 ms image acquisition and 320 µs readout time. Multiple neutrons were acquired in each pixel of these 1 ms-wide images. In that experiment the detector was synchronized to the steam engine rotation speed of 600 rpm, or 10 Hz, resulting in an acquisition speed of 750 frames/s. It was determined that 1 ms shutter was a good compromise between image blurring caused by engine operation and the neutron statistics acquired in the image of a single phase. The cylinder of a small model steam engine was installed ~6 cm from the detector active area. The trigger required for the synchronization between the piston and detector was picked up by an optical sensor installed at the flywheel connected to the piston. The preliminary imaging experiments indicated that due to a very simple and not controlled steam pressure the engine operation was quite irregular even within one cycle. Therefore, for this proof-of-principle experiment we decided to drive and control the piston by an externally coupled stepper motor. That also allowed the measurement of not only engine with the steam, but also an empty engine (no steam introduced into cylinder, only ambient air). The latter allowed the measurement of the absolute value of water content by separate calibration of the neutron attenuation by the dry engine structure. The steam engine had several leaks and allowed only ~500 s integration with the steam, after which the experiment had to be stopped, the boiler had to be cooled and refilled. Therefore the high flux of ANTARES beamline was very important for this experiment. The 9 mm aperture (resulting in L/D of ~800) and both Be (50 mm thick) and Pb (10 mm thick) filters were installed upstream in the beam. The Be filter was used to suppress the intensity of neutrons with wavelengths below ~4A in order to increase sensitivity of our imaging to the water content. In case of steam imaging even for the thickest part of the cylinder cavity (~9 mm) the attenuation of the steam at operating pressure is below our sensitivity. However, the condensation within the engine and liquid water at the leaky seals we observed in our engine can still be visualized, as demonstrated in Section **Error! Reference source not found.**

## 3. Results

### 3.1 Spectra of ANTARES beamline

The dependence of the neutron flux on neutron energy was measured first with no filters installed in the beam. During this measurement half of the detector was imaging unobscured neutrons from the source ("open beam area"), while the other half of the detector was registering neutrons which penetrated a ~10 mm BCC steel bar. Thus both the spectrum of the neutron beam itself and the transmission of a steel sample were measured simultaneously. In order to acquire good statistics per each ~10 µs-wide time slice the experiment was performed overnight. As mentioned earlier the neutron flux was substantially reduced by the beam chopper eliminating more than 99% of neutrons below 1 eV energies. As a result the largest fraction of the measured particles consisted of "unchopped" background events (epithermal and fast neutrons with E > 1 eV as well as gamma photons), see insert in Figure 1. No optimization of the background level was performed at this time and we believe a substantial reduction of the background can be achieved with proper shielding and filtering of the incoming flux if necessary. From this measurement with the chopper, where 90.9% of registered events were from the background and 9.1% from the thermal and cold neutrons we can conclude that in normal imaging, when there is no chopper installed the fraction of background events for our current detection system will be on the scale of 1-2% only, a very small fraction of detected events. After correction for the background flux the neutron spectrum measured for the open beam area is shown in Figure 1 together with the spectrum behind the steel sample, where distinct Bragg edges are clearly seen. However, one more correction has to be performed in order to reconstruct the true spectrum of the ANTARES beamline, which is described in next section.

### 3.1.1 Correction of measured spectra by the detection efficiency

The data shown in Figure 1 measured with the time of flight technique quantifies the number of registered neutrons $N_{detected}(E)$ per given neutron energy per second. That value is proportional to both the incoming neutron flux at the energy E and the detection efficiency of the device used to count the neutrons.

$$N_{detected}(E) = N_{beam}(E)*QE(E) \qquad (1)$$

Thus the energy dependence of detection efficiency $QE(E)$ needs to be taken into account for the reconstruction of true beam spectrum. In our experiments we only measure the relative variation of beam intensity with energy, rather than the absolute value (which changes with different beam collimation and filters installed in the beam) and thus the relative variation of detection efficiency is sufficient for our current analysis. For the MCP/Timepix detector the detection efficiency is defined by 3 factors $QE=P_1*P_2*P_3$ [17], where $P_1$ is the probability to absorb a neutron inside the MCP glass, $P_2$ is the probability the reaction products create a secondary particle within the pore and $P_3$ is the probability that the secondary particle creates an avalanche of electrons detected later by the Timepix readout. Our detailed modeling of detection efficiency [17], [18] as well as experimental results [19], [20] are in good agreement and indicate that probability $P_1$ has the strongest dependence on the neutron energy. $P_2$ does not change with the neutron energy and $P_3$ has some variation with E: the higher energy neutrons are absorbed closer to the MCP output and thus lead to a lower amplification. However, in our detection system optimized for neutron counting, where only one MCP is used for neutron conversion while the other MCP amplifies the event, this effect is much smaller than variation of $P_1$ and therefore we assume that $P_1$ is the dominant factor in QE variation with neutron energy and thus $QE(E) \sim Const*P_1(E)$. Disregarding interaction of neutrons with the remaining glass matrix (much smaller cross sections) the probability of neutron absorption in the MCP can be calculated from the neutron absorption cross section

$$P_1(E) = 1 - exp(-N_{10B}\sigma(E)L_{eff}) \qquad (2)$$

where $N_{10B}$ is the number of Boron atoms, $\sigma(E)$ is the neutron cross section and $L_{eff}$ is the effective thickness of MCP taking into account the open area. In the range of energies of interest from 1 meV to 1 eV the neutron cross section of 10B is known to be proportional to $1/v$ or $1/\sqrt{E}$, Figure 2. Thus probability $P_1$ can be calculated from

$$P_1(E) = 1 - \exp\left(-N_{10B}\sigma_0\sqrt{\frac{E_0}{E}}L_{eff}\right) = 1 - \exp\left(\sqrt{\frac{E_0}{E}}ln[1-P_1(E_0)]\right) \quad (3)$$

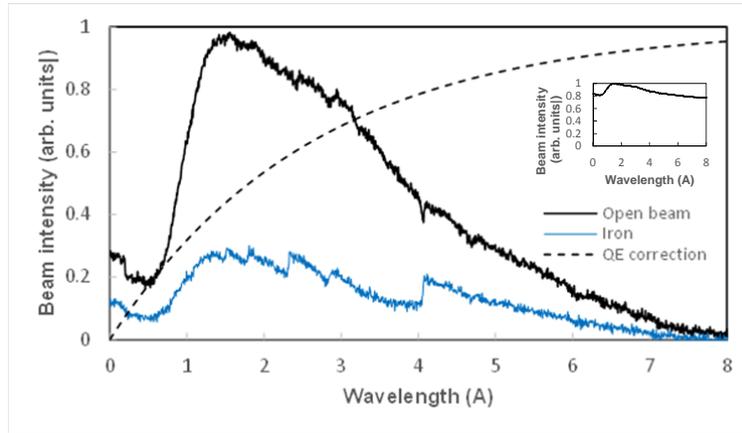

**Figure 1.** The raw spectrum of ANTARES beamline and the spectrum behind a ~10 mm bcc steel measured with a Time Of Flight technique with a disk chopper. The spectra are corrected for the background signal and not corrected for the detection efficiency. The dashed curve shows the correction function which is used for the correction of measured spectra to take into account the variation of the detector response with neutron wavelength. The insert shows the raw measured spectra of the beamline, which has a large pedestal due to the unchopped flux of fast/epithermal neutrons and gammas.

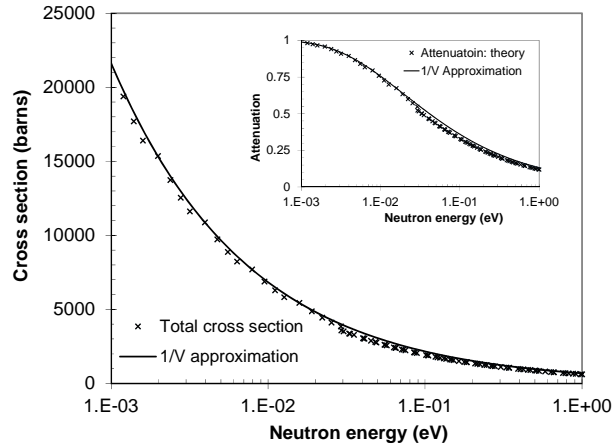

**Figure 2.** Tabulated variation of 10B total cross section (crosses) with the neutron energy and $1/V$ approximation (solid line). The insert shows the calculated attenuation of the neutron sensitive Microchannel Plate installed in the detector used in the measurements of beamline spectrum.

Equation (3) was used to calculate the relative variation of the detection efficiency with neutron energy and the beam spectrum $N_{beam}(E)$ was calculated from equation (1) with the accuracy of a multiplicative constant. The $P_1(E_0)=0.5$ value for the $E_0$ of 25 meV was taken from the results of our modeling and experimental measurement of detection efficiency [17]-[20]. The resulting correction function *Const\*QE(E)* is shown by a dashed line in Figure 1.

Figures 3-5 show the corrected spectra of the ANTARES beamline for the three beam configurations: open beam with no filters installed, Figure 3, spectrum of the neutron flux with both Be and Pb filters in the beam, Figure 4 and only with a Pb filter in the beam, Figure 5. The Be filter (which reduces the contribution of the thermal neutron flux) and the Pb filter (which can be used to reduce the relative intensity of gammas photons) introduce some spectra features due to Bragg scattering in the filters. The Be filter has distinct edges at 3.96, 3.59 and 3.47 Å, while the Pb filter seems to have a highly textured structure and introduces several dips in neutron flux at 5.23 and 3.93 Å. The simultaneously measured transmission spectrum of a BCC steel sample (~10 mm thickness) is shown in Figure 6. The steel was covering one half of the active area and therefore its measured transmission is independent of the detection efficiency which is normalized out by the division of the measured spectrum behind the sample by the spectrum in the open beam area. Several Bragg edges of the steel sample are clearly seen in the measured spectra, as well as some gradual increase of transmission before the 110 edge, indicating the presence of some texture in the Fe sample. The width of the 110 edge (which can be affected by the presence of texture in the sample) can be used to estimate the upper limit on the achievable energy resolution of our setup: the measured width of the edge (bottom of the edge to the top) is ~20 mÅ, which corresponds to ~0.5% energy resolution at 4 Å wavelength.

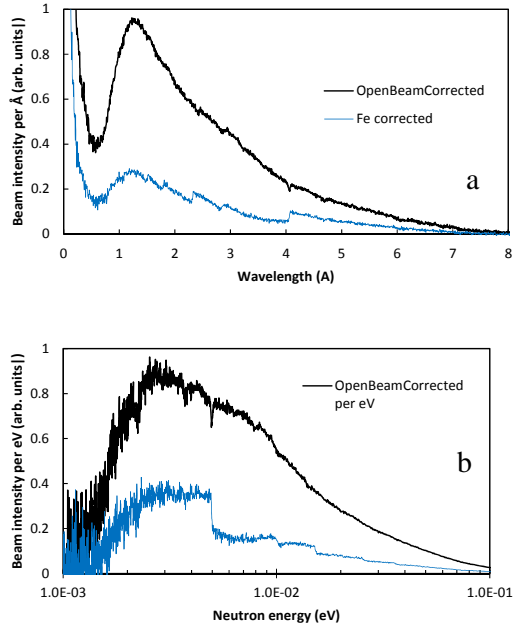

**Figure 3.** The spectrum of ANTARES beamline and the spectrum measured behind a ~10 mm steel sample, corrected for the detection efficiency. The spectrum as a function of wavelength is plotted as number of neutrons per fixed bin width of neutron wavelength $d\lambda$, while the spectrum as a function of energy is plotted as number of neutrons per fixed bin width of neutron energy $dE$. The peak of intensity is measured to be at ~1.25 Å at Figure 3.a. The contribution of fast and epithermal neutrons can be reduced by several filter options available at the ANTARES beamline.

## 3.2 Dynamics of water distribution within a steam engine

The steam engine with a cylinder of ~9 mm inner diameter, shown in Figure 7, was imaged while it was driven by an external stepper motor at 600 rotations per minute (see video of this in the online supplemental materials). Two areas of the steam engine, shown in Figure 7 by dashed rectangles, were imaged consecutively as our detector active area does not cover the entire cylinder and valves. Although we did not expect to quantify the distribution of steam in that engine as it is below our sensitivity limit, it was quickly seen that water condensed in several places inside the cylinder and remains there even after the engine was stopped and cooled down, Figure 8. The colder part of the cylinder front as well as some areas of the intake channels kept water in a condensed form continuously through the engine operation. Division of images taken with steam by the corresponding images of the same phase acquired in a dry mode eliminates the contribution of the engine structure and the resulting images have contrast due to the water/steam only, as shown in Figure 9. The condensation on the cylinder walls just after the steam was no longer supplied to the cylinder is visible in that image. It should be noted that the piston was still running during the acquisition of that image and swiping that cylinder area 10 times a second. Therefore the condensation must have been on the external sides of the cylinder, behind the metal cover. The dynamic radiography of both dry and wet cylinder operation, as well as dynamics of water only within the cylinder with all 75 frames can been seen in the online supplemental materials.

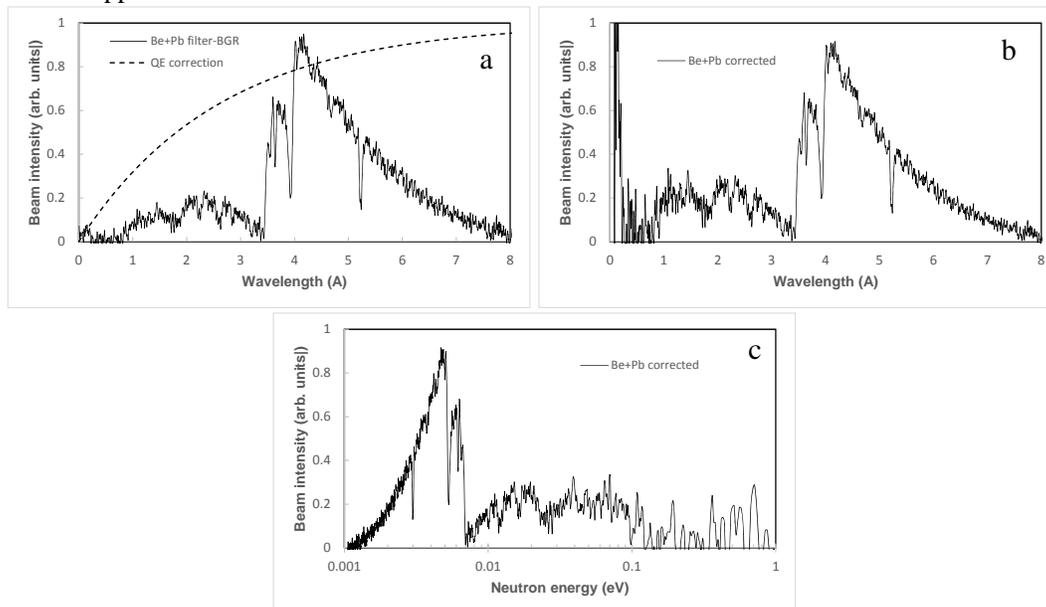

**Figure 4.** The spectrum of ANTARES beamline with Beryllium (50 mm thick) and Lead (10 mm thick) filters. (a) measured spectrum uncorrected for the detection efficiency, (b) and (c) corrected for the detection efficiency. The intensity of neutrons below 3.47 Å is substantially reduced by the filters, which can be used in experiments where colder neutron spectrum is beneficial. The Bragg edges of Be are seen at 3.47 Å (101) and 3.96 Å (100), as well as some sharp drops in intensity caused by the neutron diffraction by the strongly textured Pb filter.

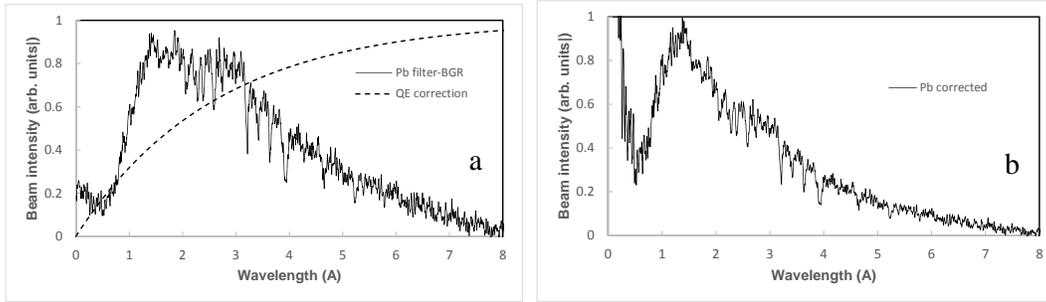

**Figure 5.** The uncorrected (a) and corrected for detection efficiency (b) spectrum of the ANTARES beamline with a Lead filter (10 mm thick) installed in the beam.

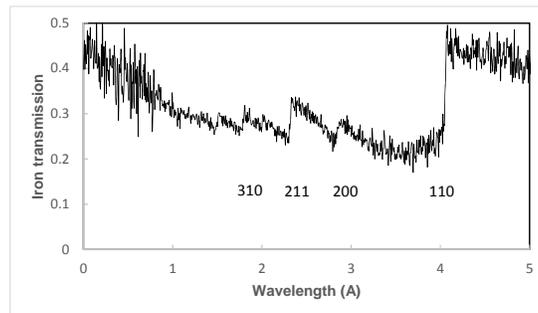

**Figure 6.** The measured transmission of a ~10 mm BCC steel sample. The 20.6 µs time binning was used during the experiment, corresponding to ~5.5 mÅ at 4 Å wavelength.

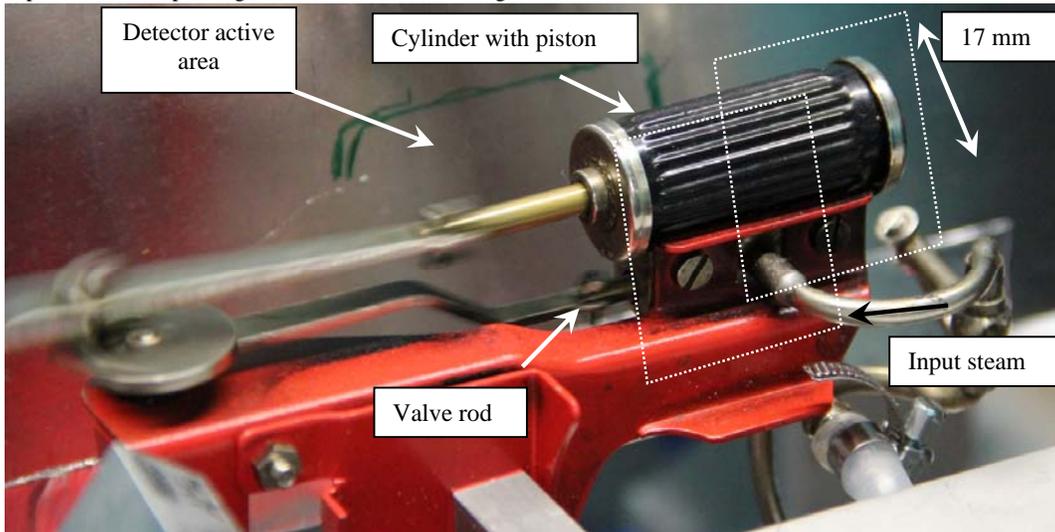

**Figure 7.** The photograph of the cylinder of a model steam engine operating at 10 Hz frequency. The dashed boxes indicate the two areas imaged in our experiments.

The snapshots of 9 frames out of 75 acquired per engine cycle is shown in Figure 10, while the sequence of 75 frames can be viewed in the supplemental online material. The steam condensation can be seen behind the piston in Figures 10.f and 10.g. It is also visible that a drop of water is formed in each cycle at the bottom left side of the engine at the point where the valve rod enters the engine. The piston position as a function of time within the cycle is shown in Figure 11, indicating the timing of the expansion and compression phases for the left side of the engine cylinder. Normalizing the images taken with steam in the engine by the dry images, phase by phase, allows visualization of water dynamics, while the rest of the structure is removed from the images, as shown in Figure 12 and the corresponding 75-frame animation in the supplementary material. The wet phase acquisition was only 500 s, which means each 1 ms-wide frame of Figs 10 and 12 has an effective acquisition time of only 5 seconds. It is the high brightness of the ANTARES beamline and high detection efficiency which allow for sufficient statistics to be acquired in such a short time, although the beam intensity can be still increased by at least an order of magnitude if a larger collimation aperture is used and filters are removed from the beam, however at the cost of spatial resolution due to the beam divergence. Combining all timing phases together improves the statistics, as shown in Figure 13, where the time-dependence of the water distribution is completely eliminated and only the average value over the entire cycle is visualized.

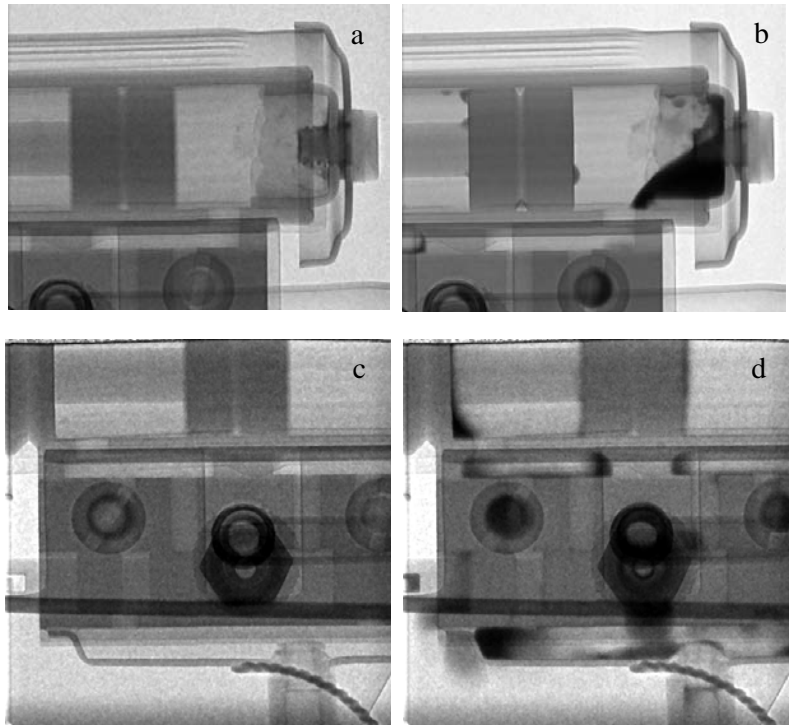

**Figure 8.** Neutron transmission radiography of the steam engine cylinder sections shown by the dashed squares in Figure 7. (a),(c) – dry engine driven by the exterior motor with no steam; (b),(d) – steam is supplied to the cylinder, engine driven by an external motor at 10 Hz; (a),(c),(d)– 1 ms-wide slices from the ~100 ms cycle (engine running at 10 Hz), some blurring of the piston is seen in the images; Only one phase is shown, while all other phases were imaged simultaneously. (b) – static engine with the remaining condensed water in it. The dry images (c), (d) were acquired over ~30 minutes period (integrated over 18000 cycles), the wet images are taken over ~17 minutes. The effective image integration time (a) – 18 sec, (b) – 100 sec, (c) – 18 sec, (d) – 10 sec.

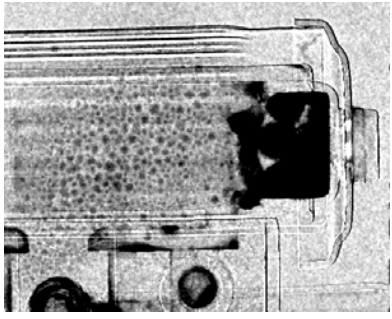

**Figure 9.** Condensation of water vapor within the steam engine cylinder measured during ~3 minutes after the steam supply was cut off. The transmission image acquired with steam condensation is normalized by the image of a dry cylinder to remove the contrast not related to vapor condensation. Some engine structure is still seen due to a small misplacement between the dry and wet measurements.

### 3.3 Time-resolved quantification of water content inside the steam engine

Not only images of water distribution can be obtained from the time-resolved images. As mentioned earlier, the normalization by the results of dry experiments allows for quantitative analysis of the effective water thickness within the engine measured as a function of time. Indeed, the flux measured by the detector when steam is introduced into the engine can be found from

$$I_{wet}(x,y,t) = I_0(x,y)e^{-\mu_{H_2O}L_{H_2O}(x,y,t)-\sum \mu_i L_i(x,y,t)} \qquad (4)$$

where $\mu_{H_2O}$ and $L_{H_2O}(x,y,t)$ are the attenuation coefficient and effective thickness of water at point *(x,y)* and time t and $\sum \mu_i L_i(x,y,t)$ are the same values for the components of the engine structure. Normalization by measurement with no steam

$$I_{dry}(x,y,t) = I_0(x,y)e^{-\sum \mu_i L_i(x,y,t)} \qquad (5)$$

eliminates the contribution from the engine components and allows quantification of effective water thickness

$$L_{H_2O}(x,y,t) = \frac{1}{\mu_{H_2O}} ln\left[\frac{I_{dry}(x,y,t)}{I_{wet}(x,y,t)}\right] \qquad (6)$$

providing the attenuation coefficient of water $\mu_{H_2O}$ is known for the given beam spectra.

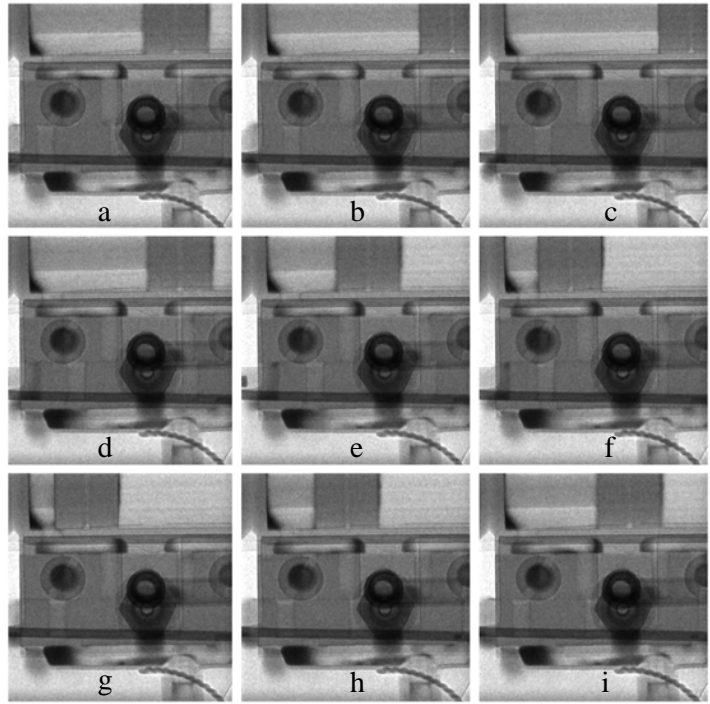

**Figure 10.** 1 ms-wide time slices of the engine operation at 10 Hz. Each slice is summed over ~10000 cycles (1000 sec integration time), corresponding to 10 sec effective image acquisition time per slice. All phases of the operation were acquired simultaneously. The position of the piston for each slice is marked by letters (a)-(i) in Figure 11.

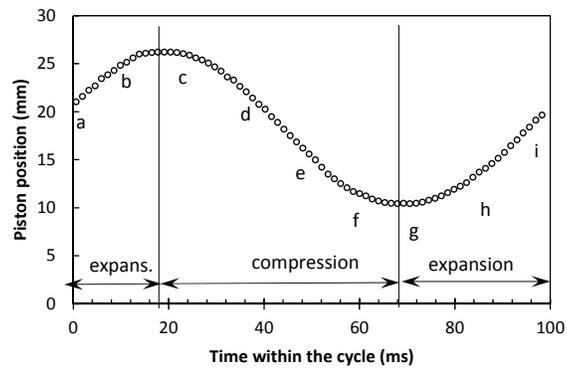

**Figure 11.** The measured position of the piston center within the cylinder.

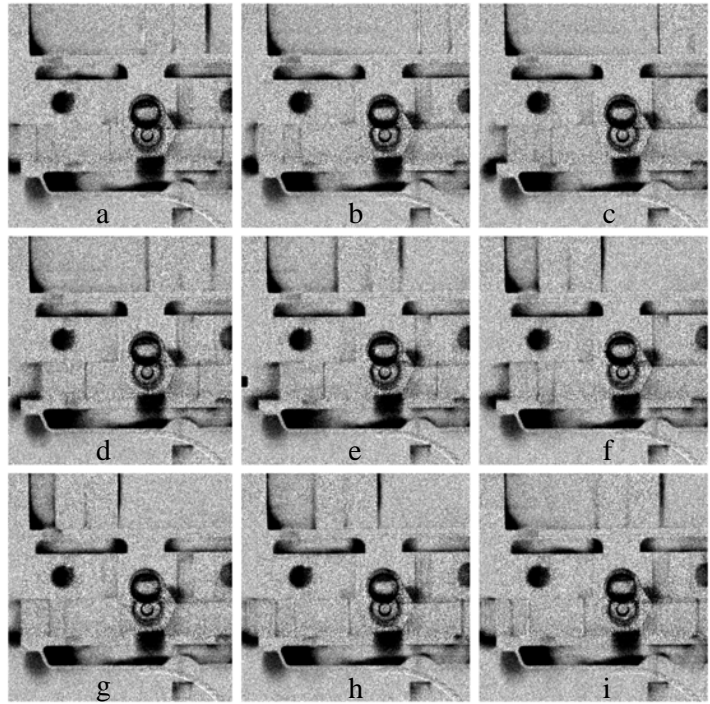

**Figure 12.** Images of running cylinder acquired with the steam normalized by the images of dry cylinder operation. The contrast in the images is caused only by the steam/water which is **repeatedly** present at the same location in the cylinder at a certain phase of the cycle. Some water condensation is always present at the surface of the piston, intake/exhaust pipes. The leak at the valve rod (bottom left corner of the image) formed a water droplet on each cycle. Phases are the same as in Figures 10 and 11. The full dynamic radiography set can be seen in the online supplemental materials.

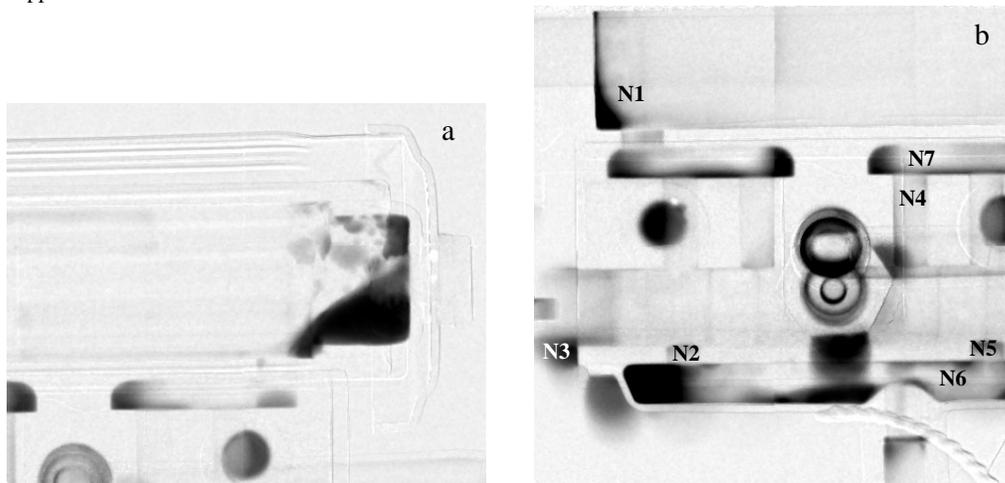

**Figure 13.** The neutron transmission image of the running cylinder with steam averaged over all phases normalized by the averaged image of the dry cylinder operation. The only contrast is due to the presence of steam/water in the cylinder. The numbers in image (b) indicate the areas for which the time dependence of the water concentration is shown in Figures 14 and 15.

Although we have not measured the attenuation coefficient of water for the ANTARES beamline, our previous experiments in a similar beam spectrum indicate that for the ANTARES beamline with the Be filter $\mu_{H_2O}$~0.3 mm-1 [21], [22]. With that assumption the time dependence of the water thickness in several areas of the steam engine was measured as shown in Figures 14 and 15. The transmission of area N1 (indicated in image 13.b) measured as a function of time is shown in Figure 14.a for both dry and wet acquisitions. The effective thickness of water calculated for that area from equation (6) is shown in Figure 14.b. There is some water with an effective thickness of ~1.4 mm remaining throughout the entire cycle in that area, and ~0.5 mm of water appears there at the end of the compression cycle and these 0.5 mm disappear by the end of the expansion cycle. The same analysis for the areas N2-N5 is shown in Figure 15. The largest amplitude in water thickness was observed at the back end of the engine where a drop of water was forming at each cycle near the valve rod entrance seal. Obviously our experiment cannot pick up the non-repetitive fluctuation of water content and only reveals the repetitive fluctuations of the water content within the steam engine.

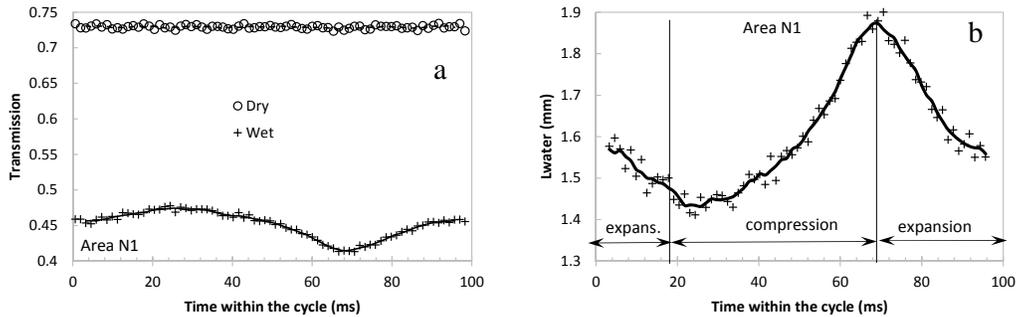

**Figure 14.** (a) - the time resolved transmission of area N1 (close to the back wall of the cylinder) measured for the dry and wet cylinder operation. (b) – measured effective thickness of water in area N1 measured as a function of time within the operation cycle.

## 4. Conclusions

The results of our experiments on time-resolved imaging at the ANTARES beamline demonstrate the unique possibilities of neutron imaging where high intensity neutron flux allows for studies of dynamic processes. High energy resolution imaging with a short pulse chopper can be an attractive alternative to monochromator- or velocity selector-based imaging in experiments where a broad range of energies has to be measured. Our fast neutron counting detector provides the possibility to acquire all energies at the same time and thus can tolerate some fluctuations of neutron flux. Indeed, in time of flight measurements all energies are acquired in simultaneously and thus can tolerate the variation of incoming neutron flux which affects all energies in the same way. Therefore the images acquired for various energies do not have to be normalized individually by the fluctuation of the neutron flux or integrated dose. Opposite to that, experiments involving scanning through neutron energies require normalization of measured spectra by the integrated neutron flux, which can vary with acquisition time, efficiency of the monochromator or velocity selector and other factors. The energy binning can also be adjusted in post-experiment processing and is not fixed by the energy width of the monochromator [23] or a velocity selector [24]. The spectrum of the newly rebuilt ANTARES beamline was measured in the 1 meV to 1 eV range in our experiments and the background signal of fast and epithermal neutrons as well as gammas was determined to be on the level of 1-2%.

The dynamic imaging of water content with a ~1 ms timing resolution and effective integration time of only 5 seconds per image of a given phase of the engine operation also demonstrates the possibility to study relatively fast repetitive processes. As demonstrated previously the timing resolution can be improved by ~2 orders of magnitude if needed [14], but the integration time in that case will have to be increased proportionally.

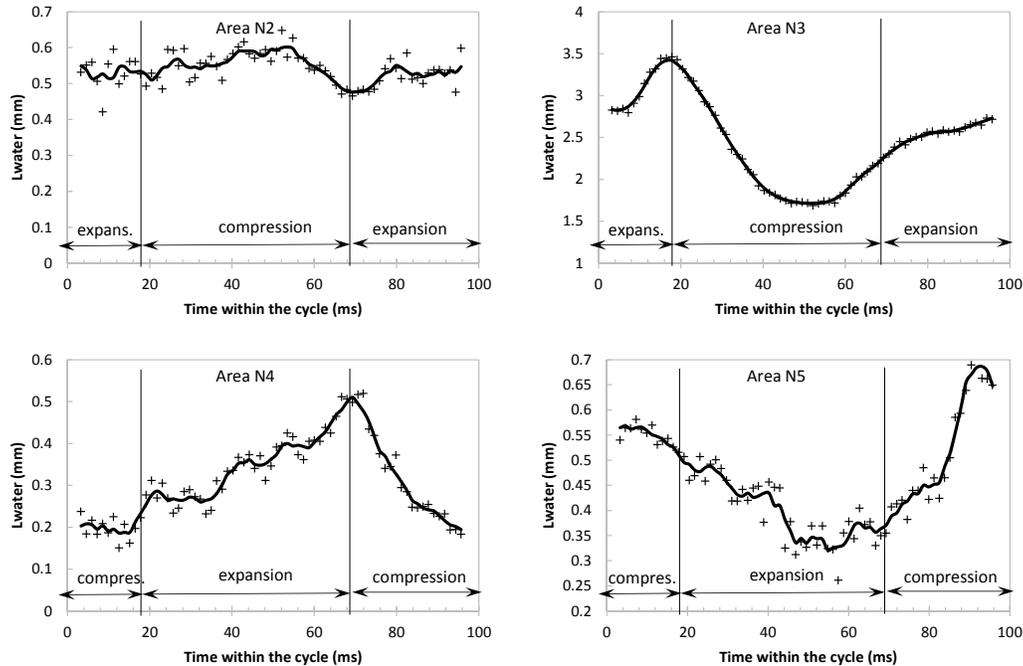

**Figure 15.** The effective water thickness measured in areas N2-N5 (defined in Figure 13) as a function of time within the operation cycle.

## Acknowledgments

We would like to acknowledge the financial support of BaCaTech foundation (BACATEC grant #28 [2012-2] "High Resolution Neutron Radiography/Tomography using Microchannel Plate based Detectors") for these experiments and also generous donation of the Vertex 5 FPGA (XC5VSX95T-1FFG1136C) Vivado design suite and DK-K7-CONN-G connectivity kit by Xilinx Inc. of San Jose, California through their Xilinx University Program. The detector used in these experiments was developed within the Medipix collaboration. This work was supported in part by U.S. Department of Energy under STTR Grants No. DE-FG02-07ER86322, DE-FG02-08ER86353 and DE-SC0009657.